\documentclass[aps,prb,twocolumn,superscriptaddress,floatfix,longbibliography]{revtex4-2}

\usepackage{amsmath,amssymb} % math symbols
\usepackage{bm} % bold math font
\usepackage{graphicx} % for figures
\usepackage{comment} % allows block comments
\usepackage{textcomp} % This package is just to give the text quote '
\usepackage{blindtext}
\usepackage{enumitem}

\setlist{noitemsep,leftmargin=*,topsep=0pt,parsep=0pt}

\usepackage{xcolor} % \textcolor{red}{text} will be red for notes
\definecolor{lightgray}{gray}{0.6}
\definecolor{medgray}{gray}{0.4}
\definecolor{myGreen}{RGB}{45, 119, 0}

\usepackage{hyperref}
\hypersetup{
colorlinks=true,
urlcolor= blue,
citecolor=blue,
linkcolor= blue,
}

\newif\ifptitle
\newif\ifpnumber
\newcounter{para}

\newcommand{\mytitle}{Probing the isolated vector magnetic field of structured laser beams by atoms}

\begin{document}

\title{\mytitle}

\author{R.~P.~Schmidt}
\email[]{riaan.schmidt@ptb.de}
\affiliation{Physikalisch-Technische Bundesanstalt, Bundesallee 100, D-38116 Braunschweig, Germany}
\affiliation{Institut für Mathematische Physik, Technische Universität Braunschweig, Mendelssohnstrasse 3, D-38106 Braunschweig, Germany}

\author{S.~{Martín-Domene}}
\affiliation{Grupo de Investigación en Aplicaciones del Láser y Fotónica, Departamento de Física Aplicada, Universidad de Salamanca, E-37008 Salamanca, Spain}
\affiliation{Unidad de Excelencia en Luz y Materia Estructuradas (LUMES), Universidad de Salamanca, E-37008 Salamanca, Spain}

\author{A.~A.~Peshkov}
\affiliation{Physikalisch-Technische Bundesanstalt, Bundesallee 100, D-38116 Braunschweig, Germany}
\affiliation{Institut für Mathematische Physik, Technische Universität Braunschweig, Mendelssohnstrasse 3, D-38106 Braunschweig, Germany}

\author{C.~{Hernández-García}}
\affiliation{Grupo de Investigación en Aplicaciones del Láser y Fotónica, Departamento de Física Aplicada, Universidad de Salamanca, E-37008 Salamanca, Spain}
\affiliation{Unidad de Excelencia en Luz y Materia Estructuradas (LUMES), Universidad de Salamanca, E-37008 Salamanca, Spain}

\author{A.~Surzhykov}
\affiliation{Physikalisch-Technische Bundesanstalt, Bundesallee 100, D-38116 Braunschweig, Germany}
\affiliation{Institut für Mathematische Physik, Technische Universität Braunschweig, Mendelssohnstrasse 3, D-38106 Braunschweig, Germany}
\affiliation{Laboratory for Emerging Nanometrology Braunschweig, Langer Kamp 6a/b, D-38106 Braunschweig, Germany}

\date{\today}

\begin{abstract}
Electric and magnetic fields are inherently coupled in an electromagnetic wave. However, structured light beams enable their spatial separation. In particular, azimuthally polarized laser beams exhibit a localized magnetic field on-axis without the electric counterpart. Recent study by Martín-Domene~\textit{et al.}~[\href{https://doi.org/10.1063/5.0197085}{App. Phys. Lett. \textbf{124}, 211101 (2024)}] has shown that combining these beams enables the generation of locally isolated magnetic fields with a controllable direction and phase. In the present paper we propose a method to probe and characterize such magnetic fields by studying their interaction with a single trapped atom. In order to theoretically investigate magnetic sublevel populations and their dependence on the relative orientation and phase---i.e. the \textit{polarization state}---of the isolated magnetic field, we use a time-dependent density-matrix method based on the Liouville–von Neumann equation. As illustrative cases, we consider the $2s^2 2p^2 \, {}^3P_0 \, - \, 2s^2 2p^2 \, {}^3P_1$, the $1s^2 2s^2 \, {}^1S_0 \, - \, 1s^2 2s 2p \, {}^3P_2$, and the $2 s^2 2p \, {}^2 P_{1/2} \, - \, 2 s^2 2p \, {}^2 P_{3/2}$ transitions in ${}^{40}$Ca$^{14+}$, ${}^{10}$Be, and ${}^{38}$Ar$^{13+}$, respectively. Our results indicate that monitoring atomic populations serves as an effective tool for probing isolated vector magnetic fields, which opens avenues for studying laser-induced processes in atomic systems where electric field suppression is critical.
\end{abstract}

\maketitle
\section{\label{sec:Introduction}Introduction}

Structured light is an emerging cutting-edge field that refers to the ability to shape different degrees of freedom of an electromagnetic wave such as the amplitude, phase, and polarization state~\cite{Torres:2011,BliokhPR2015,BarnettJO2016, 2023_shen_Roadmap, 2023_bliokh_Roadmap}. This field was boosted by the discovery that light beams with helical wave fronts, known as twisted modes, possess a finite projection of the orbital angular momentum (OAM)~\cite{AllenPhysRevA1992,MolinaNP2007,PadgettOE2017}, which has found many applications in optics, nanophotonics, and atomic physics~\cite{YaoAOP2011,FrankeArnoldPTRS2017,RubinszteinDunlopJO2017,MivelleNanoLett2025}.

Besides twisted modes, more complex field distributions can also be achieved. In particular, combinations of beams where the OAM and spin angular momentum (SAM) are antiparallel to each other give rise to vector beams (VB), characterized by their spatial dependent polarization state~\cite{HallDGOptLett1996}. Remarkably, in order to accomplish the Maxwell equations, VB present nonzero longitudinal components on-axis~\cite{LaxPRA1975,BliokhPRA2010,QuinteiroPRL2017}, leading to distinct light-matter interactions~\cite{QuinteiroPRA2017,BabikerJO2019,IketakiOL2007,MonteiroPRA2009,MonteiroPRA2012,PeshkovAdP2023}.

A noteworthy example of VB are the so-called azimuthally polarized beams. Such beams possess an azimuthal electric-field polarization pattern with a singularity on the vortex line, i.e. at the location of a longitudinal magnetic field~\cite{VeysiJOSAB2015}. Particular interest has been given to the enhancement of this isolated longitudinal magnetic field. Examples include the use of metallic nanoantennas~\cite{BlancoACSPhot2019,Martin-HernandezRPhotRes2024}, photoionization in gases~\cite{SederbergPhysRevX2020}, or plasmonic nanostructures \cite{Reynier2023, Reynier2025}. Once being enhanced, the range of applications that rely only on interactions with the locally isolated magnetic field is wide within different research fields such as magnetic spectroscopy~\cite{VeysiJOSAB2016,WozniakNJP2021}, force microscopy~\cite{ZengJO2023}, optical spectroscopy~\cite{KasperczykPRL2015}, ultrafast nonlinear magnetization dynamics in ferromagnets~\cite{SanchezHPLSciEng2023}, and chiral media~\cite{TangPRL2010,CheonNPJQuantMat2022}. 

A recent work proposed to superpose multiple azimuthally polarized VB propagating in different directions in order to generate locally isolated magnetic fields with a controllable direction and phase, i.e. isolated optical magnetic fields with controllable polarization state~\cite{MartinDomeneAPL2024}. The magnetic field is isolated in the sense that the electric field vanishes at the point where the beam axes intersect, making the magnetic field ``isolated'' at that point. Up to the present, however, no experimental scheme exists to determine the orientation of such an isolated magnetic field, which we refer to as ``magnetic light'' in line with the term widely used in the fields of nanophotonics and plasmonics~\cite{KuznetsovSciRep2012}. In the present work, we propose a method to characterize the magnetic-light orientation by studying its interaction with a single atom. Our approach relies on the analysis of population dynamics of ground-state magnetic sublevels. We demonstrate that population distribution strongly depends on the polarization state of ``magnetic light'', and hence can be used to characterize this polarization.

This paper is organized as follows. In Sec.~\ref{sec:Theory} we briefly recall the basic formulas needed to describe ``magnetic light'' as superpositions of azimuthally polarized VB. Moreover, the interaction of this light with an atom is treated by introducing Rabi frequency and density-matrix formalism based on the Liouville–von Neumann equation. Calculations for such Rabi frequencies have been performed for the $2s^2 2p^2 \, {}^3P_0 \, - \, 2s^2 2p^2 \, {}^3P_1$ magnetic dipole ($M1$) transition in Ca$^{14+}$ in Sec.~\ref{subsec:Calcium} and the $1s^2 2s^2 \, {}^1S_0 \, - \, 1s^2 2s 2p \, {}^3P_2$ magnetic quadrupole ($M2$) transition in Be in Sec.~\ref{subsec:Beryllium}. Based on these calculations we found that atomic transitions induced by the ``magnetic light'' exhibit different selection rules than those known from photoexcitation by plane-waves. Detailed analysis of time-dependent atomic population dynamics is presented in Sec.~\ref{subsec:Argon} for the $2 s^2 2p \, {}^2 P_{1/2} \, - \, 2 s^2 2p \, {}^2 P_{3/2}$ transition in Ar$^{13+}$. Our findings indicate that the orientation of the isolated magnetic field strongly affects magnetic sublevel populations. Finally, a summary of our results and an outlook are given in Sec.~\ref{sec:Summary}.

\section{\label{sec:Theory}Theory}

\subsection{\label{subsec:TheoryLight}Structured light modes}

In this work, we investigate the coupling of a target atom to an isolated optical magnetic field. Below, we will show that this ``magnetic light'' can be constructed from a superposition of azimuthally polarized VB. The basic properties of such beams will be briefly recalled in Sec.~\ref{subsubsec:AzPolLight}, while their superposition will be discussed in Sec.~\ref{subsubsec:MagneticLight}.

\subsubsection{\label{subsubsec:AzPolLight}Azimuthally polarized light beams}

\begin{figure*}[t]
	\centering
	\includegraphics[width=0.8\textwidth]{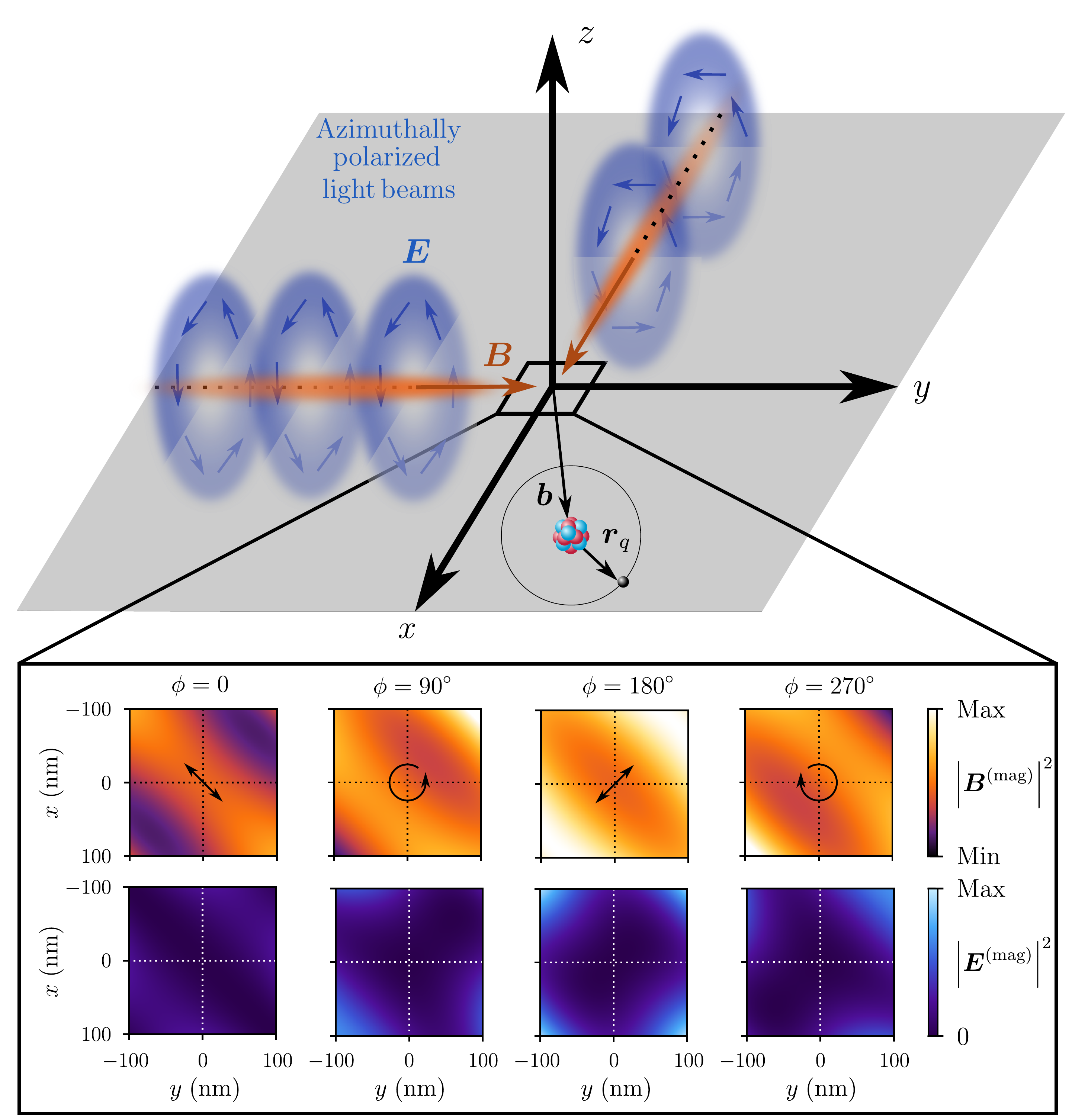}
	\caption{Geometry for the excitation of a single atom by a superposition of two azimuthally polarized VB, which propagate in the laboratory frame $x$ and $y$ direction, respectively. The intersection point of the beam axes is chosen as the origin of the coordinate system. The position of a target atom with respect to the origin is characterized by the impact parameter $\boldsymbol{b}$. The quantization $(z)$ axis is perpendicular to the propagation directions of both beams. The bottom inset shows the squared magnetic (top row) and electric (bottom row) fields given by Eqs.~\eqref{eq:CombinedEField}~and~\eqref{eq:CombinedBField} for weights $c_1=c_2=1/\sqrt{2}$, opening angle $\theta_k = 0.54^\circ$, and photon energy $\hbar \omega = 2.73$ eV corresponding to a wavelength of $455$ nm. The arrows indicate the polarization of the magnetic field at the origin. The electromagnetic field is shown only in the vicinity of the intersection point, since the target atom is localized there.}
	\label{fig:Geometry}
\end{figure*}

Azimuthally polarized light beams are typically described in terms of Laguerre-Gaussian or Hermite-Gaussian modes with different polarization states. In this work, we choose to construct the azimuthally polarized beam in the basis of the well-known Bessel states~\cite{JentschuraPRL2011}, as it allows a simpler treatment of structured-light matter interaction. These states are characterized by the longitudinal linear momentum $k_z$, the absolute value of the transverse momentum $|\boldsymbol{k}_\perp| = \varkappa$, the frequency $\omega = c \sqrt{k_z^2 + \varkappa^2}=ck$, the helicity $\lambda$, and the well-defined projection $m_\gamma$ of the total angular momentum onto the propagation direction. The vector potential of Bessel radiation can be written in cylindrical coordinates ($r_\perp^\prime,\phi_r^\prime,z^\prime$) in the form

\begin{equation}
\begin{split}
    \boldsymbol{A}^\mathrm{(B)}_{m_\gamma , \lambda} (\boldsymbol{r}^\prime,t) =& \, A_0 \sum\limits_{m_s = 0,\pm 1} \boldsymbol{e}_{m_s} (-i)^{m_s} c_{m_s} J_{m_\gamma - m_s}(\varkappa r_\perp^\prime) \\
    &\times \mathrm{exp}[i(m_\gamma - m_s) \phi_r^\prime + i k_z z^\prime - i \omega t] ,
\end{split}
\label{eq:VP_TW}
\end{equation}

\noindent where the prime indicates that the $z^\prime$ axis, chosen along the light propagation direction, does not necessarily coincide with the quantization $z$ axis of the overall system. In Eq.~\eqref{eq:VP_TW}, $A_0$ denotes the field amplitude, $J_{m_\gamma - m_s}(\varkappa r_\perp^\prime)$ is the Bessel function of the first kind, and $\boldsymbol{e}_{\pm 1} = (\boldsymbol{e}_{x^\prime} \pm i \boldsymbol{e}_{y^\prime}) / \sqrt{2}$ and $\boldsymbol{e}_0 = \boldsymbol{e}_{z^\prime}$ are the polarization vectors. Moreover, the coefficients $c_{m_s}$ read as $c_{\pm 1} = \pm \lambda (1 \pm \lambda \cos\theta_k)/2$ and $c_0 = -\sin\theta_k /\sqrt{2}$, where $\theta_k = \mathrm{arctan}(\varkappa / k_z)$ is the polar opening angle~\cite{Matula/JPB:2013,SerboUsp2018}.

An azimuthally polarized VB can be written as a linear combination of two Bessel
beams with $\lambda = \pm 1$ and $m_\gamma = 0$:

\begin{equation}
\begin{split}
    \boldsymbol{A}^\mathrm{(az)} (\boldsymbol{r}^\prime,t) = \frac{1}{\sqrt{2}} \Bigl[& \boldsymbol{A}^\mathrm{(B)}_{m_\gamma=0 , \lambda=-1} (\boldsymbol{r}^\prime,t) \\
    &- \boldsymbol{A}^\mathrm{(B)}_{m_\gamma=0 , \lambda=+1} (\boldsymbol{r}^\prime,t) \Bigr].
\end{split}
\label{eq:VP_TW_Azim}
\end{equation}

\noindent See Ref.~\cite{Schulz/PRA:2020} for more information. To better understand the properties of $\boldsymbol{A}^\mathrm{(az)}$, we expand it in multipoles:

\begin{equation}
    \boldsymbol{A}^\mathrm{(az)} (\boldsymbol{r}^\prime) = \sqrt{4\pi} \sum_L i^L \sqrt{2L+1} \; d^L_{0,-1} (\theta_k) \; \boldsymbol{a}^{(0)}_{L0}(\boldsymbol{r}^\prime),
\label{eq:MultipolesAzPolLight}
\end{equation}

\noindent where $d^L_{0,1}$ are elements of the small Wigner $d$-matrix~\cite{Rose:1957}. The complete absence of electric multipoles $\boldsymbol{a}^{(1)}_{L0}$ in Eq.~\eqref{eq:MultipolesAzPolLight} means that the target atom placed on the beam axis ``sees'' only magnetic multipoles $\boldsymbol{a}^{(0)}_{L0}$ of rank $L$. However, when the atom is displaced from the beam axis, both magnetic and electric multipoles contribute to the light-atom interaction.

Let us now calculate electric and magnetic fields of the azimuthally polarized VB. By using the standard relations $\boldsymbol{E}=-\partial_t \boldsymbol{A}$ and $\boldsymbol{B}=\nabla \times \boldsymbol{A}$, we obtain:

\begin{subequations}
\begin{align}
    \boldsymbol{E}^\mathrm{(az)} (\boldsymbol{r}^\prime,t) =& \, i \omega A_0 J_1(\varkappa r_\perp^\prime) \mathrm{exp}[i ( k_z z^\prime - \omega t )] \boldsymbol{e}_{\phi_r^\prime} , \label{eq:E_Field_Azim} \\
    \boldsymbol{B}^\mathrm{(az)} (\boldsymbol{r}^\prime,t) =& -i k A_0 \cos\theta_k J_1(\varkappa r_\perp^\prime) \mathrm{exp}[i ( k_z z^\prime - \omega t )] \boldsymbol{e}_{r_\perp^\prime} \notag \\
    &+ k A_0 \sin\theta_k J_0(\varkappa r_\perp^\prime) \mathrm{exp}[i ( k_z z^\prime - \omega t )] \boldsymbol{e}_{z^\prime} , 
\end{align}
\label{eq:E_and_B_Field_Azim}
\end{subequations}

\noindent where $\boldsymbol{e}_{r_\perp^\prime}$, $\boldsymbol{e}_{\phi_r^\prime}$ and $\boldsymbol{e}_{z^\prime}$ are the basis unit vectors. From Eq.~\eqref{eq:E_Field_Azim}, we see that the electric field vector is directed along the azimuthal direction $\boldsymbol{e}_{\phi_r^\prime}$, thus justifying the name \textit{azimuthally polarized} beam. For the present work, it is important to note that the electric field $\boldsymbol{E}^\mathrm{(az)}$ vanishes on the beam axis, $r_\perp^\prime = 0$, while the longitudinal component of the magnetic field is nonzero:

\begin{subequations}
\begin{align}
    \boldsymbol{E}^\mathrm{(az)} (r_\perp^\prime = 0, z^\prime,t) =& \, 0, \\
    \boldsymbol{B}^\mathrm{(az)} (r_\perp^\prime = 0, z^\prime,t) =& \, k A_0 \sin\theta_k \mathrm{exp}[i ( k_z z^\prime - \omega t )] \boldsymbol{e}_{z^\prime} ,
\end{align}
\label{eq:E_and_B_Field_Azim_BeamAxis}
\end{subequations}

\noindent as discussed in Ref.~\cite{QuinteiroPRA2017}.

\subsubsection{\label{subsubsec:MagneticLight}Magnetic light}

At the beam center, $r_\perp^\prime = 0$, an azimuthally polarized VB~\eqref{eq:E_and_B_Field_Azim_BeamAxis} exhibits a longitudinal magnetic field along the beam axis, $\boldsymbol{B}^\mathrm{(az)} \parallel \boldsymbol{e}_{z^\prime}$. By applying a second azimuthally polarized beam, one can rotate $\boldsymbol{B}$ in arbitrary direction~\cite{MartinDomeneAPL2024}. In Fig.~\ref{fig:Geometry}, for example, we consider a geometry where two azimuthally polarized VB propagate perpendicular to each other along the $x$ and $y$ axes. In this chosen geometry the quantization $(z)$ axis is directed perpendicular to the axes of both beams. The reference frame defined in this way will be referred to as the \textit{laboratory frame} with the origin at the intersection of both beam axes.

For the geometry displayed in Fig.~\ref{fig:Geometry}, the vector potential of a superposition of two azimuthally polarized VB takes the form

\begin{equation}
    \boldsymbol{A}^\mathrm{(mag)} (\boldsymbol{r},t) = c_\mathrm{1} \boldsymbol{A}^\mathrm{(az)}_x (\boldsymbol{r},t) + c_\mathrm{2} \, \mathrm{exp}[i \phi]  \boldsymbol{A}^\mathrm{(az)}_y (\boldsymbol{r},t) .
\label{eq:VP_Admixture}
\end{equation}

\noindent Here the real weights satisfy $c_\mathrm{1}^2 + c_\mathrm{2}^2 = 1$, and $\phi$ is the relative phase between the components. It should be noted that in Eq.~\eqref{eq:VP_Admixture} both vector potentials $\boldsymbol{A}^\mathrm{(az)}_x$ and $\boldsymbol{A}^\mathrm{(az)}_y$ are given in the laboratory frame, whereas in Sec.~\ref{subsubsec:AzPolLight} they are written in the local frame with the $z^\prime$ axis along the light propagation direction. The transformation from one reference frame to another can be readily done and the details are omitted.

Similar to Eq.~\eqref{eq:MultipolesAzPolLight}, we expand the light field in Eq.~\eqref{eq:VP_Admixture} in multipoles:

\begin{equation}
\begin{split}
    \boldsymbol{A}^\mathrm{(mag)}(\boldsymbol{r}) =& \sqrt{4 \pi} \sum_{LM} i^L \sqrt{2L+1} \; d^L_{0,-1}(\theta_k) \; d^L_{M,0}(\pi/2) \\
    &\times \; \boldsymbol{a}^{(0)}_{LM} (\boldsymbol{r})  \Bigl( c_1  + c_2 \, \mathrm{exp}[i(\phi - M \, \pi/2)] \Bigr).
\end{split}
\label{eq:MultipolesMagneticLight}
\end{equation}

\noindent A conclusion from Eq.~\eqref{eq:MultipolesMagneticLight} is that an atom placed at the origin, $\boldsymbol{r}=0$, can interact only with magnetic multipoles. For this reason, we will refer to the field described by Eq.~\eqref{eq:VP_Admixture} as ``magnetic light''. As we have already mentioned, the exclusion of electric multipoles from the coupling between light and matter depends on the position of the atom, but not on its size. Thus the concept of ``magnetic light'' is determined not only by the properties of the light itself, but also by the position of the observer (atom).

By making use of the vector potential~\eqref{eq:VP_Admixture}, one can obtain the electric and magnetic fields of the superposed beams. The expressions for these fields are given in Appendix~\ref{app:EandBfields} for arbitrary $\boldsymbol{r}$, and their time-averaged spatial distributions, $\left| \boldsymbol{E}^\mathrm{(mag)} \right|^2$ and $\left| \boldsymbol{B}^\mathrm{(mag)} \right|^2$, are presented in the bottom inset of Fig.~\ref{fig:Geometry} for different phases $\phi$. As seen from this figure, ``magnetic light'', contrary to its name, exhibits not only a magnetic, but also an electric field. This is not surprising, because magnetic multipoles of the radiation field consist of both magnetic and electric fields~\cite{Rose:1957,Akhiezer:1965}. But despite the presence of the electric field, ``magnetic light'' can only excite magnetic multipole transitions if an atom is located at $\boldsymbol{r}=0$. As we will see below, the excitation process reflects the symmetry of ``magnetic light'' at the origin. The electric and magnetic fields at this point are:

\begin{subequations}
\begin{align}
    \boldsymbol{E}^\mathrm{(mag)}(\boldsymbol{r}=0,t) =& \, 0, \label{eq:EFieldatZero} \\
    \boldsymbol{B}^\mathrm{(mag)}(\boldsymbol{r}=0,t) =& \, k A_0 \sin\theta_k \mathrm{exp}[-i\omega t] \notag \\
    &\times  \Bigl(c_1\boldsymbol{e}_x + c_2 \mathrm{exp}[i\phi] \boldsymbol{e}_y \Bigr). \label{eq:BFieldatZero}
\end{align}
\label{eq:CombinedEandBFieldatZero}
\end{subequations}

\noindent As seen from these expressions, the electric field $\boldsymbol{E}^\mathrm{(mag)}$ vanishes at $\boldsymbol{r}=0$, in contrast to the nonzero magnetic field $\boldsymbol{B}^\mathrm{(mag)}$, which is directed perpendicular to the quantization axis. Such a magnetic field oscillates at optical frequency and its ``polarization'' is determined by the phase $\phi$ and the coefficients $c_1$ and $c_2$. It should be recalled that polarization usually refers to the orientation of the \textit{electric field} of an electromagnetic wave. Since $\boldsymbol{E}^\mathrm{(mag)}(\boldsymbol{r}=0) = 0$, we will exploit this term to specify the direction of the \textit{magnetic field} at the origin. In the case when $c_1=c_2=1/\sqrt{2}$, the phase $\phi=0^\circ$ or $180^\circ$ leads to linear polarization, while $\phi=90^\circ$ or $270^\circ$ corresponds to circular polarization. See Ref.~\cite{MartinDomeneAPL2024} for more details.

\subsection{\label{subsec:RabiFrequency}Rabi frequency}

\subsubsection{\label{subsubsec:RabiFrequencyPlaneWave}Plane-wave radiation}

We continue our theoretical analysis with a brief reminder of the excitation of an atom by plane-wave radiation. In particular, we consider a transition between ground $\left|\alpha_g F_g M_g \right>$ and excited $\left|\alpha_e F_e M_e \right>$ atomic states with the total angular momentum $\boldsymbol{F} = \boldsymbol{I} + \boldsymbol{J}$, where $\boldsymbol{I}$ and $\boldsymbol{J}$ are the nuclear and electron angular momenta, respectively, $M$ is the projection of $\boldsymbol{F}$ onto the $z$ axis, and $\alpha$ refers to all additional quantum numbers. We assume that this transition proceeds via only one multipole channel $(pL)$ and that the plane wave propagates along the $z$ axis and has polarization $\boldsymbol{e}^\mathrm{(pl)}= c_2 \, \mathrm{exp}[i \phi] \boldsymbol{e}_{x} - c_1 \boldsymbol{e}_{y}$. Such a parametrization ensures that the direction of the magnetic field of the plane wave coincides with the local magnetic field of the ``magnetic light'' at $\boldsymbol{r}=0$, see Eq.~\eqref{eq:BFieldatZero}. With these assumptions the Rabi frequency, which characterizes the coupling between the two atomic states, can be written as

\begin{equation}
\begin{split}
    \Omega_R^\mathrm{(pl)} =& \, \Biggl| \frac{A_0 e c}{\hbar} \Biggl< \alpha_e F_e M_e \bigg| \sum\limits_q  \boldsymbol{\alpha}_q \cdot \boldsymbol{e}^\mathrm{(pl)} e^{i \boldsymbol{k} \cdot \boldsymbol{r}_q} \bigg| \alpha_g F_g M_g \Biggr> \Biggr| \\
    =& \; \widetilde{\Omega}_R \biggl| \frac{i^p}{\sqrt{2}} \left< F_g \, M_g \, L \, 1 | F_e \, M_e \right> (i c_1 + c_2 \mathrm{exp}[i \phi]) \\
    &\; - \frac{(-i)^p}{\sqrt{2}} \left< F_g \, M_g \, L \, -1 | F_e \, M_e \right> (i c_1 - c_2 \mathrm{exp}[i \phi]) \biggr| ,
\end{split}
\label{eq:RabiFrequencyPlaneWave}
\end{equation}

\noindent where we have introduced the so-called reduced frequency

\begin{equation}
\begin{split}
    \widetilde{\Omega}_R =& \, \biggl| \frac{A_0 e c}{\hbar} \sqrt{2 \pi (2 L + 1) (2 F_g +1)} \begin{Bmatrix} 
      F_e & F_g & L \\ 
      J_g & J_e & I 
   \end{Bmatrix} \\
   &\times \left< \alpha_e J_e || H_\gamma (pL) || \alpha_g J_g \right> \biggr| ,
\label{eq:ReducedRabiFrequency}
\end{split}
\end{equation}

\noindent and $\left< \alpha_e J_e || H_\gamma (pL) || \alpha_g J_g \right>$ denotes the reduced matrix element for the magnetic ($p=0$) or electric ($p=1$) transition of the order $L$~\cite{GorenPRA2004}. Moreover, $\boldsymbol{r}_q$ determines the position of the $q$th electron with respect to the atomic center of mass, and $\boldsymbol{\alpha}_q$ denotes the vector of Dirac matrices~\cite{Johnson:2007,Wense:2020}.

The Rabi frequency~\eqref{eq:RabiFrequencyPlaneWave} is nonzero if the angular-momentum and parity selection rules

\begin{subequations}
\begin{align}
    |F_e - F_g| \leq& \ L \leq F_e + F_g , \\[0.2cm]
    M_e - M_g =& \ \Delta M = \pm 1 , \label{eq:SelectionRulesPlaneWavesDeltaM} \\[0.2cm]
    \pi_g \pi_e =& \ (-1)^{L+p+1} ,
\end{align}
\label{eq:SelectionRulesPlaneWaves}
\end{subequations}

\noindent are satisfied. Here, $\pi_g$ and $\pi_e$ are the parities of the ground and excited states, respectively~\cite{Rose:1957}.

\subsubsection{\label{subsubsec:RabiFrequencyMagneticLight}Magnetic light}

Having examined the interaction with plane-wave radiation, we move on to a discussion of the coupling between a target atom and the ``magnetic light''. Let us, for the moment, assume that the center of mass of the atom is located at the origin, $\boldsymbol{r}=0$. As discussed above, at this position only magnetic multipoles contribute to the excitation, see Eq.~\eqref{eq:MultipolesMagneticLight}.

It can be shown after some angular momentum algebra that the Rabi frequency for the transition between the $\left|\alpha_g F_g M_g \right>$ and $\left|\alpha_e F_e M_e \right>$ states induced by the ``magnetic light''~\eqref{eq:VP_Admixture} is given by

\begin{equation}
\begin{split}
    \Omega_R^\mathrm{(mag)} =& \, \left| \frac{e c}{\hbar} \left< \alpha_e F_e M_e \bigg| \sum\limits_q  \boldsymbol{\alpha}_q \cdot \boldsymbol{A}^\mathrm{(mag)} (\boldsymbol{r}_q) \bigg| \alpha_g F_g M_g \right> \right| \\
    =& \; \widetilde{\Omega}_R \Bigl| \left< F_g \, M_g \, L \, \Delta M | F_e \, M_e \right> \\
    &\qquad \times \sqrt{2} \; \delta_{0,p} \; d^L_{0,1}(\theta_k) \; d^L_{\Delta M,0}(\pi/2) \\
    &\qquad \times \Bigl( c_1  + c_2 \, \mathrm{exp}[i(\phi - \Delta M \, \pi/2)] \Bigr) \Bigr| ,
\end{split}
\label{eq:RabiFrequencyMagneticLight}
\end{equation}

\noindent From Eq.~\eqref{eq:RabiFrequencyMagneticLight} we deduce the selection rules

\begin{subequations}
\begin{align}
    |F_e - F_g| \leq& \; L \leq F_e + F_g , \\[0.2cm]
    M_e - M_g = \Delta M =& \, -L , -L+2 , \dots , L-2 , L , \label{eq:SelectionRulesMagneticLightDeltaM} \\[0.2cm]
    p=0 \; \; \mathrm{and}& \; \; \pi_g \pi_e = \ (-1)^{L+1} .
\end{align}
\label{eq:SelectionRulesMagneticLight}
\end{subequations}

\noindent Note that the $\Delta M$ selection rule~\eqref{eq:SelectionRulesMagneticLightDeltaM} is in general different from that for plane waves~\eqref{eq:SelectionRulesPlaneWavesDeltaM}. Indeed, the difference of angular momentum projection $\Delta M$ changes in steps of two. This angular-momentum selection rule is related to the fact that $d^L_{\Delta M,0}(\pi /2) \propto \delta_{L-\Delta M , 2n}$, where $n$ is an integer~\cite{1988_Varshalovich}. There is one unique case where Eqs.~\eqref{eq:SelectionRulesMagneticLight} coincide with the plane-wave selection rules given by Eqs.~\eqref{eq:SelectionRulesPlaneWaves}. This is the $M1$ transition for which $F_e - F_g = 0, \pm 1$, $\Delta M = \pm 1$, and $\pi_g \pi_e = 1$. In the case of non-dipole transitions, Eqs.~\eqref{eq:SelectionRulesMagneticLight} predict a different selection rule for $\Delta M$. In particular, for an $M2$ transition we have $\Delta M = 0, \pm 2$, in contrast to $\Delta M = \pm 1$ for the plane-wave case.

\subsection{\label{subsec:DensityMatrix}Density matrix formalism}

The Rabi frequency provides general information about the strength of the coupling between a target atom and the incident radiation. In the absence of relaxation, this frequency uniquely determines the time dependence of populations of atomic states. In order to investigate the time evolution of the system in the presence of relaxation, a more detailed analysis based on the density matrix theory is required~\cite{2010_Budker,Blum:2012}. In this approach, the state of the system is represented by the density operator $\hat{\rho}(t)$ that satisfies the Liouville–von Neumann equation:

\begin{equation}
    \frac{\mathrm{d}}{\mathrm{d} t}\hat{\rho}(t) = -\frac{i}{\hbar} \left[ \hat{H}(t), \hat{\rho}(t) \right] + \hat{R}(t) .
\label{eq:Liouville_Original}
\end{equation}

\noindent Here, $\hat{H}(t)$ is the total Hamiltonian of an atom in the presence of both the incident radiation and an external magnetic field. Moreover, the operator $\hat{R}(t)$ has been added ``by hand'' to account for relaxation due to radiative decay~\cite{Tremblay/PRA:1990}.

For further analysis, it is more convenient to rewrite Eq.~\eqref{eq:Liouville_Original} in matrix form. For this purpose, we choose a set of basis states $\left| \alpha F M \right>$. The resulting system of coupled differential equations for the density-matrix elements can be found in Refs.~\cite{Schmidt/PRA:2024,Ramakrishna/PRA:2024}. We then solve these equations to find the density matrix as a function of time.

In order to visualize the results and simplify the discussion, it is practical to describe the population of atomic states in terms of the so-called statistical tensors

\begin{equation}
\begin{split}
    \rho_{kq} (\alpha F ; t) =& \, \sum\limits_{M \, M^\prime} (-1)^{F-M^\prime} \left< F \, M \, F \, -M^\prime | k \, q \right> \\
    &\times \left< \alpha F M | \hat{\rho} (t) | \alpha F M^\prime \right>,
\end{split}
\label{eq:StatisticalTensors}
\end{equation}

\noindent which transform like spherical harmonics of rank $k$ under a rotation of the coordinates~\cite{2000_Balashov}. The normalization of $\rho_{kq}$ by means of the zero-rank tensor

\begin{equation}
    \mathcal{A}_{kq} (\alpha F ; t) = \frac{\rho_{kq} (\alpha F ; t)}{\rho_{00} (\alpha F ; t)}
\label{eq:OrientationAndAlignmentParameters}
\end{equation}

\noindent gives rise to the orientation ($k$ odd) and alignment ($k$ even) parameters. These parameters describe the relative population of atomic sublevels $\left| \alpha F M \right>$ and coherence between them.

\section{\label{sec:Results}Results and Discussion}

Equations~\eqref{eq:RabiFrequencyPlaneWave}-\eqref{eq:OrientationAndAlignmentParameters} can be applied to any atom or ion, independent of its nuclear spin and electronic shell structure. Below, we consider three systems: (i) the $2s^2 2p^2 \; {}^3P_0 \, - \, 2s^2 2p^2 \; {}^3P_1$ transition in ${}^{40}$Ca$^{14+}$, (ii) the $1s^2 2s^2 \; {}^1S_0 \, - \, 1s^2 2s 2p \; {}^3P_2$ transition in ${}^{10}$Be, and (iii) the $2 s^2 2p \; {}^2 P_{1/2} - 2 s^2 2p \; {}^2 P_{3/2}$ transition in ${}^{38}$Ar$^{13+}$. In all cases, the nuclear spin is $I=0$, so that we are dealing with transitions between fine-structure levels. The atoms are placed directly at the intersection point of both beam axes. The VB are tuned to exact resonance with the transition. It is assumed that $c_1 = c_2 = 1/\sqrt{2}$. Such weights lead to linear polarization of light when $\phi=0^\circ$ or $180^\circ$ and circular polarization when $\phi=90^\circ$ or $270^\circ$. Moreover, we have chosen the parameters $A_0$ and $\theta_k$ such that the radius of the first ring of the Bessel beam is equal to $31$ wavelengths and the peak value of the electric field on this ring is $1.7 \times 10^{-13} \, \omega \, \mathrm{V/m}$. With these conditions, the isolated magnetic field strength is of the order of $10^{-8}$~T. Experiments with such a small isolated magnetic field will require very good shielding from external magnetic fields. At the same time, static (at the atomic scale) magnetic fields, such as the Earth's magnetic field, do not induce atomic transitions, but can only cause a Zeeman shift. For this reason, external static magnetic fields do not interfere with the optically isolated magnetic field in atomic transitions.

\subsection{\label{subsec:Calcium}${}^\textbf{3}$\textbf{\textit{P}}${}_\textbf{0} \rightarrow {}^\textbf{3}$\textbf{\textit{P}}${}_\textbf{1}$ transition in Ca$^{14+}$}

\begin{figure}[t]
	\centering
	\includegraphics[width=0.45\textwidth]{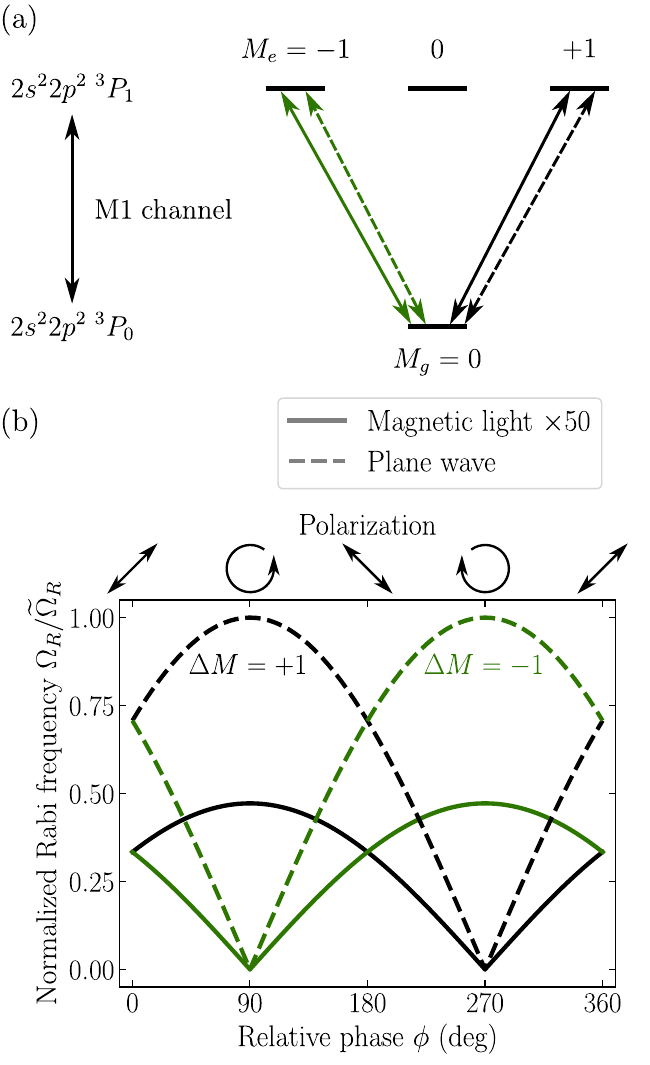}
	\caption{(a) The $2s^2 2p^2 \, {}^3P_0 \, - \, 2s^2 2p^2 \, {}^3P_1$ magnetic dipole transition in ${}^{40}$Ca$^{14+}$. The arrows indicate the interaction with ``magnetic light'' (solid) and plane waves (dashed). Shown are $M_g=0 \rightarrow M_e=+1$ (black) and $M_g=0 \rightarrow M_e=-1$ (green) transitions. The atom is located at the origin of the coordinate system. (b) Normalized Rabi frequencies for these transitions as a function of the relative phase $\phi$ which determines the light polarization (top). Here, the weights are $c_1 = c_2 = 1/\sqrt{2}$, the photon energy is $\hbar \omega = 2.18$ eV, and the polar opening angle is $\theta_k = 0.54^\circ$. The values for ``magnetic light'' are increased by a factor of 50.}
	\label{fig:CaClikeRabiFrequencies}
\end{figure}

We start with the ${}^3P_0 - {}^3P_1$ $M1$ transition in ${}^{40}$Ca$^{14+}$ ion. The relatively long lifetime of the excited state of about $10.5$~ms~\cite{NaqviThesis1951} allows us to neglect spontaneous decay for interaction times shorter than 1~ms. At such time intervals, the entire evolution of the system is simply Rabi oscillations at the frequencies $\Omega_R$. Fig.~\ref{fig:CaClikeRabiFrequencies} shows graphs of $\Omega_R$ divided by the reduced frequency~\eqref{eq:ReducedRabiFrequency}. The latter is equal to $\widetilde{\Omega}_R = 2\pi \times 215$~kHz for the chosen set of parameters. The results for plane waves (dashed lines) and ``magnetic light'' (solid lines) are presented as a function of the relative phase $\phi$, which determines the polarization. As seen from Fig.~\ref{fig:CaClikeRabiFrequencies}, the Rabi frequencies for both light fields show qualitatively the same behavior. Namely, only the $M_e = \pm 1$ magnetic sublevels are excited, as follows from the selection rules~\eqref{eq:SelectionRulesPlaneWaves}~and~\eqref{eq:SelectionRulesMagneticLight}, and $\Omega_R^\mathrm{(mag)}$ has exactly the same phase dependence as $\Omega_R^\mathrm{(pl)}$. In particular, for $\phi=0^\circ$ the magnetic field of both types of radiation has linear polarization, leading to $\Omega_R(M_e=+1)=\Omega_R(M_e=-1)$. In contrast, $\Omega_R(M_e=+1) \neq 0$ and $\Omega_R(M_e=-1) = 0$ for $\phi=90^\circ$, which corresponds to magnetic field with circular polarization.

\subsection{\label{subsec:Beryllium}${}^\textbf{1}$\textbf{\textit{S}}${}_\textbf{0} \rightarrow {}^\textbf{3}$\textbf{\textit{P}}${}_\textbf{2}$ transition in Be}

\begin{figure}[t]
	\centering
	\includegraphics[width=0.45\textwidth]{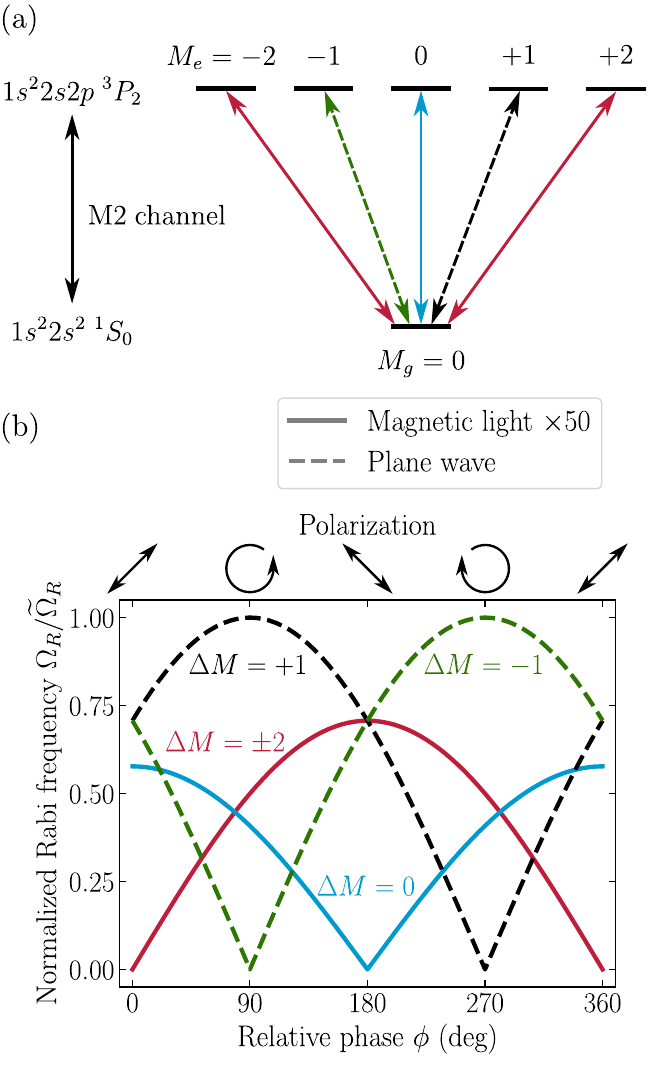}
	\caption{(a) The $1s^2 2s^2 \, {}^1S_0 \, - \, 1s^2 2s 2p \, {}^3P_2$ magnetic quadrupole transition in ${}^{10}$Be. ``Magnetic light'' induces  $M_g=0 \rightarrow M_e=\pm 2$ (red solid line) and $M_g=0 \rightarrow M_e=0$ (blue solid line) transitions. Plane waves induce $M_g=0 \rightarrow M_e=+1$ (black dashed line) and $M_g=0 \rightarrow M_e=-1$ (green dashed line) transitions. (b) Phase dependence of normalized Rabi frequencies. The photon energy is $\hbar \omega = 2.73$ eV. All other parameters are the same as in Fig.~\ref{fig:CaClikeRabiFrequencies}.}
	\label{fig:BeRabiFrequencies}
\end{figure}

In contrast to the ${}^3P_0 - {}^3P_1$ $M1$ case described above, the ${}^1S_0 - {}^3P_2$ transition in ${}^{10}$Be proceeds via the $M2$ channel. Since the multipole is higher, the transition is weaker. As a consequence, the lifetime of the excited state increases to 6061~s~\cite{TachievJoPB1999} and the reduced frequency decreases to $\widetilde{\Omega}_R = 2\pi \times 327$~Hz. Fig.~\ref{fig:BeRabiFrequencies} shows the transition scheme and the normalized Rabi frequencies $\Omega_R^\mathrm{(pl)}/\widetilde{\Omega}_R$ and $\Omega_R^\mathrm{(mag)}/\widetilde{\Omega}_R$ as a function of $\phi$. It can be seen that plane waves and ``magnetic light'' drive different transitions. The plane wave only couples the $M_g = 0$ ground state to the $M_e = \pm 1$ excited state, while the ``magnetic light'' drives transitions $\left| M_g = 0 \right> \rightarrow \left| M_e = 0, \pm 2 \right>$. Such behavior follows the selection rules given by Eqs.~\eqref{eq:SelectionRulesPlaneWaves}~and~\eqref{eq:SelectionRulesMagneticLight}, as discussed above.

In addition to different $\Delta M$ selection rules, Fig.~\ref{fig:BeRabiFrequencies} also shows qualitatively different dependence of $\Omega_R^\mathrm{(pl)}$ and $\Omega_R^\mathrm{(mag)}$ on the phase $\phi$, and hence polarization. For example, for $\phi=90^\circ$ and $270^\circ$ both plane waves and ``magnetic light'' exhibit right and left circular polarization. For the plane-wave case this circular polarization can be attributed to helicity $\pm 1$ which results in transition between sublevels with $\Delta M = \pm 1$. The predictions for the ``magnetic light'' look counterintuitive: despite of circular polarization, one can induce $\Delta M = 0, \pm 2$ transitions. This behavior can be understood from the conservation of total angular momentum projection on the quantization axis. Indeed, by analyzing the expectation value of the $z$-component of the total angular momentum operator $\left< \hat{J}_z \right>$, one obtains the well-know result $\left< \hat{J}_z \right> = \hbar \lambda$ for plane waves with circular polarization. In contrast, $\left< \hat{J}_z \right> = 0$ for $L=2$ multipoles of circularly polarized ``magnetic light''. Together with the selection rules~\eqref{eq:SelectionRulesMagneticLight}, this implies that only excitation to the sublevels $M_e = 0, \pm 2$ is possible. For the latter case, the substates $\left|M_e=\pm 2 \right>$ must be equally populated, so that $\left< \hat{J}_z^\mathrm{(atom)} \right> = 0$.

In the case when $\phi=0^\circ$ and $180^\circ$ plane waves and magnetic-light radiation are linearly polarized. For plane waves, both orientations of linear polarization result in equal population of $M_e= \pm 1$ substates, i.e. $\Omega_R^\mathrm{(pl)}(M_e = +1) = \Omega_R^\mathrm{(pl)}(M_e = -1)$. In contrast, ``magnetic light'' with opposite directions of linear polarization leads to different coupling of atomic sublevels. While, for example, $\Omega_R^\mathrm{(mag)}(M_e=0) \neq 0$ for $\phi=0^\circ$, it vanishes for $\phi=180^\circ$. To explain this behavior, we refer to Fig.~\ref{fig:Geometry} which indicates that rotation of the polarization vector of ``magnetic light'' by $90^\circ$ is connected to a change of electric and magnetic field distributions and hence the symmetry of the system.

\subsection{\label{subsec:Argon}${}^\textbf{2}$\textbf{\textit{P}}${}_\textbf{1/2} \rightarrow {}^\textbf{2}$\textbf{\textit{P}}${}_\textbf{3/2}$ transition in Ar$^{13+}$}

In the previous sections, we discussed the Rabi frequency $\Omega_R$ for $M1$ and $M2$ transitions in ${}^{40}$Ca${}^{14+}$ and ${}^{10}$Be atoms, respectively. In what follows, we consider time-dependent population dynamics for the ${}^{2}P_{1/2} \, - \, {}^{2}P_{3/2}$ transition in ${}^{38}$Ar${}^{13+}$ including relaxation due to spontaneous emission, see Fig.~\ref{fig:SystemTimeAlignmentOrientation}~(a). This spontaneous emission can proceed via $M1$ and electric quadrupole $(E2)$ channels. However, the decay rate $\Gamma_{E2}$ is rather small compared to $\Gamma_{M1}$~\cite{FischerData2012}, and hence $E2$ decay will be omitted. To investigate atomic population dynamics, we solve the Liouville-von Neumann equation~\eqref{eq:Liouville_Original}.

\subsubsection{\label{subsubsec:ComparisonPlaneWave}Orientation parameter $\mathcal{A}_{10}$}

\begin{figure}[t]
	\centering
	\includegraphics[width=0.44\textwidth]{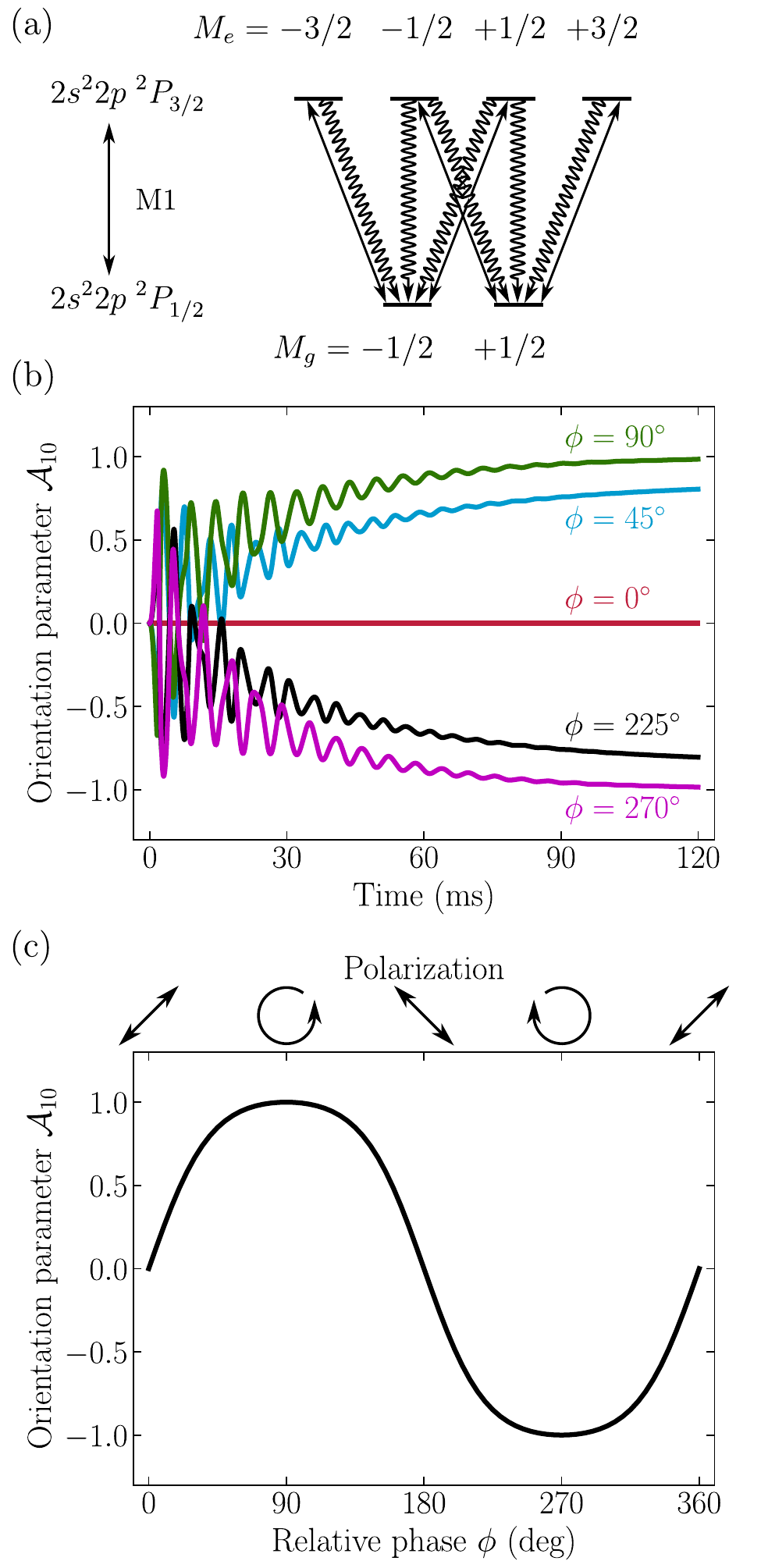}
	\caption{(a) The $2 s^2 2p \; {}^2 P_{1/2} \, - \, 2 s^2 2p \; {}^2 P_{3/2}$ magnetic dipole transition in ${}^{38}$Ar$^{13+}$. The arrows represent the interactions with ``magnetic light''~(\ref{eq:VP_Admixture}), and the wavy lines represent spontaneous decay. (b) Time dependence of the orientation $\mathcal{A}_{10}$ of the $2 s^2 2p \; {}^2 P_{1/2}$ ground state for different relative phases: $\phi = 0^\circ$ (red), $\phi = 45^\circ$ (blue), $\phi = 90^\circ$ (green), $\phi = 225^\circ$ (black), and $\phi = 270^\circ$ (magenta). (c) Steady-state orientation $\mathcal{A}_{10}$ as a function of the relative phase $\phi$. The photon energy is $\hbar \omega = 2.81$ eV. All other parameters are the same as in Fig.~\ref{fig:CaClikeRabiFrequencies}.}
	\label{fig:SystemTimeAlignmentOrientation}
\end{figure}

Below we will pay special attention to the relative population of the $\left| 2s^2 2p \; {}^{2}P_{1/2} \; M_g=\pm 1/2 \right>$ magnetic sublevels. As discussed in Sec.~\ref{subsec:DensityMatrix}, it is convenient to describe this relative population in terms of the orientation parameter

\begin{equation}
    \mathcal{A}_{10} (t) = \frac{\rho_{+1/2}(t) - \rho_{-1/2}(t)}{\rho_{+1/2}(t) + \rho_{-1/2}(t)} ,
\label{eq:A10inBlikeAr}
\end{equation}

\noindent with substate population

\begin{equation}
    \rho_{M_g}(t) = \left< 2s^2 2p \, {}^{2}P_{1/2} \, M_g \right| \hat{\rho}(t) \left| 2s^2 2p \, {}^{2}P_{1/2} \, M_g \right> ,
\label{eq:RhoMginBlikeAr}
\end{equation}

\noindent which follows from Eqs.~\eqref{eq:StatisticalTensors}~and~\eqref{eq:OrientationAndAlignmentParameters}. The time evolution of the ground-state orientation $\mathcal{A}_{10}$, produced in the course of the interaction with the ``magnetic light'', is shown in Fig.~\ref{fig:SystemTimeAlignmentOrientation}~(b). It can be seen from the figure that the atomic sublevel population reaches steady state in hundreds of milliseconds after performing several Rabi oscillations. As described in Ref.~\cite{Fox:2006}, this behavior corresponds to light damping regime for which the ratio of the damping rate $\Gamma_{M1}=0.1$~kHz to the Rabi frequency $\Omega_R^\mathrm{(mag)} \approx 1$~kHz is of the order of $\Gamma_{M1} / \Omega_R^\mathrm{(mag)} \approx 0.1$. In the following, we focus on the analysis of the steady-state orientation parameter, i.e. $\mathcal{A}_{10}(t>120~\mathrm{ms})$.

The steady-state orientation $\mathcal{A}_{10}$ is shown in Fig.~\ref{fig:SystemTimeAlignmentOrientation}~(c) as a function of the relative phase $\phi$. From this figure and Eq.~\eqref{eq:A10inBlikeAr}, one can see that ${}^{2}P_{1/2}$ ground state is fully oriented for $\phi=90^\circ$ and $270^\circ$ which corresponds to right and left circular polarization of ``magnetic light'', respectively. This resembles the outcome of photoexcitation by circularly polarized plane waves and can be understood by analysis of conservation of total angular momentum projection. Indeed, $\left< \hat{J}_z \right> = \pm\hbar$ for ``magnetic light'' with $\phi=90^\circ$ and $270^\circ$ which results in pumping of the population into the magnetic substate $M_g=\pm 1/2$ after several Rabi cycles. In contrast, in the case when $\phi=0^\circ$ and $180^\circ$, the ``magnetic light'' is linearly polarized and carries $\left< \hat{J}_z \right> = 0$, but this does not lead to atomic orientation. Such a zero atomic orientation is also well known from photoexcitation by plane waves with linear polarization. It follows from Fig.~\ref{fig:SystemTimeAlignmentOrientation}~(c) and our discussion that the orientation of the ${}^2 P_{1/2}$ ground state can be used to study the magnetic-light polarization. This orientation can be measured, for example, by state dependent fluorescence~\cite{BlattEJP1988,SchmiegelowNatCom2016}.

\subsubsection{\label{subsubsec:BeamPower}Beam-weight dependence of $\mathcal{A}_{10}$}

Above we showed that the orientation $\mathcal{A}_{10}$ of the ${}^2 P_{1/2}$ state of Ar$^{13+}$ is very sensitive to the relative phase between magnetic-light components~\eqref{eq:VP_Admixture} with equal weights $c_1 = c_2 = 1/\sqrt{2}$. In this scenario both azimuthally polarized VB have equal total power. However, the weights $c_1$ and $c_2$ may not always be equal in experiments. In order to account for such unequal beam weights and make the proposed diagnostic method even more accessible, we study here the atomic ground-state orientation for cases when $c_1 \neq c_2$.

We note that when $c_1 \neq c_2$ the phase $\phi$ still uniquely determines the handedness of the magnetic-light polarization. Together, $c_1$, $c_2$, and $\phi$ specify this
polarization which might differ from that obtained in the case of equal weights. Indeed, as seen from Eq.~\eqref{eq:BFieldatZero} the ``magnetic light'' is linearly polarized for all values of $c_1$ and $c_2$ when $\phi=0^\circ$ and $180^\circ$. For $\phi=90^\circ$ and $270^\circ$ as well as $c_1 \neq c_2$, Eq.~\eqref{eq:BFieldatZero} predicts ``magnetic light'' with elliptical polarization which approaches linear polarization when $c_1 \gg c_2$ or $c_1 \ll c_2$.

\begin{figure}[t]
	\centering
	\includegraphics[width=0.48\textwidth]{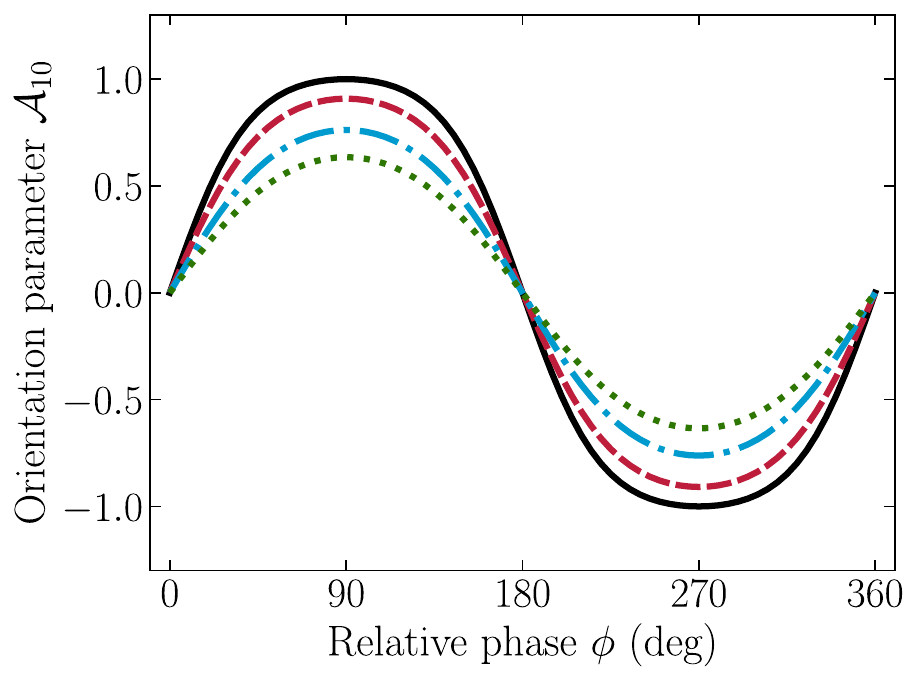}
        \caption{Same as Fig.~\ref{fig:SystemTimeAlignmentOrientation}~(c), but for the coefficients $c_\mathrm{1}=c_\mathrm{2}$ (black solid line), $c_\mathrm{1}=2c_\mathrm{2}$ (red dashed line), $c_\mathrm{1}=3c_\mathrm{2}$ (blue dash-dotted line), and $c_\mathrm{1}=4c_\mathrm{2}$ (green dotted line).}
	\label{fig:PowerDependency}
\end{figure}

The orientation $\mathcal{A}_{10}$ for different weights $c_1$ and $c_2$ is displayed in Fig.~\ref{fig:PowerDependency} as a function of $\phi$. It is shown that the relative-phase sensitivity of the atomic orientation is maximal for $c_1 = c_2$, i.e. when the ``magnetic light'' changes polarization state from linear to circular. This sensitivity becomes less pronounced in the limit $c_1 \gg c_2$ as follows from analysis of $\left< \hat{J}_z \right>$. Indeed, we find $\left< \hat{J}_z \right>=0$ for all weights when $\phi=0^\circ$ and $180^\circ$, leading to vanishing $\mathcal{A}_{10}$. In contrast, $\Bigl| \left< \hat{J}_z (c_1=c_2) \right> \Bigr|=\hbar > \Bigl| \left< \hat{J}_z (c_1 \neq c_2) \right> \Bigr|$ in the case when $\phi=90^\circ$ and $270^\circ$ which results in declining atomic orientation.

\subsubsection{\label{subsubsec:DelocalizedAtoms}Effect of delocalized atom on atomic orientation}

All calculations above have been carried out for the atomic center of mass perfectly localized at $\boldsymbol{r}=0$, which is the intersection of the beam axes of the azimuthally polarized components. However, achieving perfect localization is experimentally unrealistic owing to laser jittering and thermal motion of trapped atoms, which introduce uncertainty in the position of the atom. To account for this delocalization, we assume that the probability of finding an atom at a distance $\boldsymbol{b}$ from the beam intersection (see Fig.~\ref{fig:Geometry}) is given by

\begin{equation}
    f(\boldsymbol{b}) = \frac{1}{2 \pi \sigma^2} e^{-\frac{\boldsymbol{b}^2}{2 \sigma^2}} \, ,
\end{equation}

\noindent with a width $\sigma$ \cite{SerboUsp2018,Schulz/PRA:2020}. By using $f(\boldsymbol{b})$, one can calculate the average sublevel population

\begin{equation}
    \overline{\rho}_{M_g}(t) = \int f(\boldsymbol{b}) \, \rho_{M_g}(t) \, \mathrm{d}^3 \boldsymbol{b} \, ,
    \label{eq:AveragedRho}
\end{equation}

\noindent and the average orientation parameter $\overline{\mathcal{A}}_{10}$ by replacing $\rho_{M_g}(t)$ with $\overline{\rho}_{M_g}(t)$ in Eq.~\eqref{eq:A10inBlikeAr}. In the past this semi-classical approach has been successfully employed to describe the excitation of a trapped ion by vortex radiation~\cite{LangePRL2022}. For the sake of simplicity, we assume that the atom is delocalized only within the $x$-$y$ plane, see Fig.~\ref{fig:Geometry}.

It should be noted that the spatial spread will cause the atom to couple to the electric quadrupole component of the field~\cite{DuanJPB2019}. This leads to the fact that the ${}^2 P_{1/2} \, - \, {}^2 P_{3/2}$ transition may proceed not only via the $M1$ but also the $E2$ channel. However, for Ar${}^{13+}$ the $E2$ channel is much weaker than the $M1$ one and does not contribute significantly to atomic population dynamics.

\begin{figure}[t]
	\centering
	\includegraphics[width=0.48\textwidth]{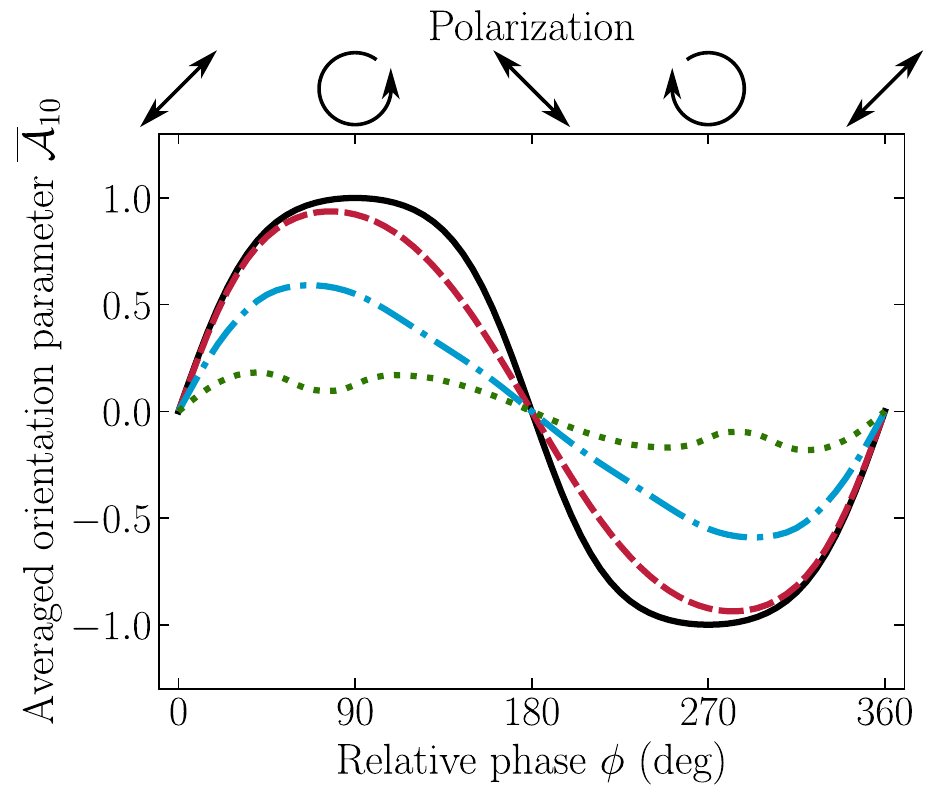}
	\caption{Same as Fig.~\ref{fig:SystemTimeAlignmentOrientation}~(c), but for different sizes of the atomic target: $\sigma=0$ (black solid line), $\sigma=20$ nm (red dashed line), $\sigma=50$ nm (blue dash-dotted line), and $\sigma=100$ nm (green dotted line).}
	\label{fig:Sigma}
\end{figure}

The average orientation $\overline{\mathcal{A}}_{10}$ is displayed in Fig.~\ref{fig:Sigma} as a function of $\phi$ for the weights $c_1 = c_2 = 1/\sqrt{2}$. In order to illustrate the atom delocalization effect, calculations have been performed for $\sigma=20$ nm, $\sigma=50$ nm, as well as $\sigma=100$ nm, and compared with the ideal case of $\sigma=0$. Target sizes of several tens of nanometers have already been demonstrated in experiments with twisted light and single trapped ions~\cite{SchmiegelowNatCom2016,LangePRL2022}. As seen from the figure, the delocalization of the target atom affects the sensitivity of the orientation parameter to the phase $\phi$. For example, $\bigl|\overline{\mathcal{A}}_{10}\bigr|$ varies between $0$ and almost $1$ for the relatively small width $\sigma = 20$~nm. In contrast, for the target size of $100$~nm, which is less than a hundredth of the radius of the VB intensity ring, the average orientation lies in the range $-0.18 \leq \overline{\mathcal{A}}_{10} \leq 0.18$. Such an effect on the atomic orientation can complicate light polarization measurements.

\section{\label{sec:Summary}Summary and Outlook}

In summary, we have performed a theoretical analysis of the excitation of a single atom by a combination of azimuthally polarized light beams exhibiting a local isolated magnetic field. Special attention has been paid to magnetic sublevel population of the atom as well as to the question how this population is affected by the relative phase between the beam components. This phase can be attributed to the polarization of the isolated magnetic field at the beam axes intersection. In order to explore the sensitivity to the polarization of such ``magnetic light'', we have solved the Liouville–von Neumann equation for the time evolution of
the atomic density matrix.

While the formalism developed here can be applied to any atom and transition, in the present study we considered the $2s^2 2p^2 \, {}^3P_0 \, - \, 2s^2 2p^2 \, {}^3P_1$, the $1s^2 2s^2 \, {}^1S_0 \, - \, 1s^2 2s 2p \, {}^3P_2$, and the $2 s^2 2p \; {}^2 P_{1/2} \, - \, 2 s^2 2p \; {}^2 P_{3/2}$ transitions in ${}^{40}$Ca$^{14+}$, ${}^{10}$Be, and ${}^{38}$Ar$^{13+}$, respectively. Based on the analysis of these transitions, we have shown that different selection rules apply to the excitation by the ``magnetic light'' compared to conventional plane-wave radiation. In particular, all electric multipole transitions are forbidden when the atomic center of mass is located at the beam axes intersection. For such a localized atom, moreover, the difference of angular momentum projection between the atomic ground and excited state changes within the multipole order in increments of two.

For the $2 s^2 2p \, {}^2 P_{1/2} \, - \, 2 s^2 2p \, {}^2 P_{3/2}$ transition in ${}^{38}$Ar$^{13+}$ we have performed detailed calculations of the atomic population dynamics due to the interaction with ``magnetic light''. Based on these calculations we found significant orientation of the ${}^2 P_{1/2}$ ground state which strongly depends on the relative phase and weights of the azimuthally polarized components. This orientation holds under experimental conditions in which the beam amplitudes differ and the atom is imprecisely localized with respect to intersection of the two beam axes. Following our theoretical results, we propose that analysis of the atomic ground-state orientation can serve as a valuable tool for diagnostics of polarization of optical isolated magnetic fields.

In order to make the proposed scheme more applicable to experiments, a broader theoretical treatment for the interaction of atoms with ``magnetic light'' is required. In particular, while we restricted our current work to continuous waves of azimuthally polarized light with well-defined phase relationship, we aim to explore the coupling of confined atoms to ultrashort pulses of Laguerre-Gaussian radiation featuring finite laser coherence time in a forthcoming study.

By adding another non-vortex laser beam, the isolated magnetic field can also be characterized using atomic ensembles. Similar to the STED technique~\cite{DrechslerPRL2021}, the non-vortex beam is tightly focused inside the vortex beam. A transition scheme is chosen such that without the first beam, the sample appears ``transparent'' to the vortex beam. In this case, the alignment of the beam axes ensures that only the atoms on the axis act as a probe for the isolated magnetic field.

\begin{acknowledgments}
This project has received funding from the Deutsche Forschungsgemeinschaft (DFG, German Research Foundation) under Project-ID 445408588 (SU 658/5-2) and Project-ID 274200144, under SFB 1227 within project B02, under Germany’s Excellence Strategy, EXC-2123 QuantumFrontiers, Project No. 390837967, from the QuantERA II Programme via the EU H2020 research and innovation programme under Grant No. 101017733, EPSRC under Grant No. EP/Z000513/1 (V-MAG), and from the European Research Council (ERC) under the European Union’s Horizon 2020 research and innovation programme (grant agreement No 851201). C.H.-G. acknowledges funding from Ministerio de Ciencia	e Innovación (Grant PID2022-142340NB-I00) and the Department of Education of the Junta de Castilla y León and FEDER Funds (Escalera de Excelencia CLU-2023-1-02 and grant No. SA108P24). The authors are grateful for fruitful discussions with M.~Naise.
\end{acknowledgments}

\appendix

\section{\label{app:EandBfields}Combined electric and magnetic fields}

From the vector potential in Eq.~\eqref{eq:VP_Admixture} and by using the relations $\boldsymbol{E}=-\partial_t \boldsymbol{A}$ and $\boldsymbol{B}=\nabla \times \boldsymbol{A}$, one can derive the electric and magnetic field of the superposition of azimuthally polarized VB for the geometry shown in Fig.~\ref{fig:Geometry}. By separating the time-dependent parts, $\boldsymbol{E}^\mathrm{(mag)}(\boldsymbol{r},t)=\boldsymbol{E}^\mathrm{(mag)}(\boldsymbol{r})\mathrm{exp}[-i \omega t]$ and $\boldsymbol{B}^\mathrm{(mag)}(\boldsymbol{r},t)=\boldsymbol{B}^\mathrm{(mag)}(\boldsymbol{r})\mathrm{exp}[-i \omega t]$, the position-dependent terms in the laboratory frame read

\begin{equation}
\begin{split}
    &\boldsymbol{E}^\mathrm{(mag)}(\boldsymbol{r}) \\
    =& \, \boldsymbol{e}_x c_2 i \omega A_0 J_1(\varkappa \sqrt{x^2+z^2})\mathrm{exp}[i k_z y + i\phi] z/\sqrt{x^2+z^2} \\
    -& \boldsymbol{e}_y c_1 i \omega A_0 J_1(\varkappa \sqrt{y^2+z^2})\mathrm{exp}[i k_z x] z/\sqrt{y^2+z^2} \\
    +& \boldsymbol{e}_z \Bigl( c_1 i \omega A_0 J_1(\varkappa \sqrt{y^2+z^2})\mathrm{exp}[i k_z x] y/\sqrt{y^2+z^2} \\
    &\; - c_2 i \omega A_0 J_1(\varkappa \sqrt{x^2+z^2})\mathrm{exp}[i k_z y + i\phi] x/\sqrt{x^2+z^2} \Bigr), \\
\end{split}
\label{eq:CombinedEField}
\end{equation}

\begin{equation}
\begin{split}
    &\boldsymbol{B}^\mathrm{(mag)}(\boldsymbol{r}) \\
    =& \, \boldsymbol{e}_x \Bigl( c_1 k A_0 \sin\theta_k J_0(\varkappa\sqrt{y^2+z^2})\mathrm{exp}[i k_z x] \\
    &\; - c_2 i k A_0 \cos\theta_k J_1(\varkappa \sqrt{x^2+z^2})\mathrm{exp}[i k_z y + i\phi] \\
    &\quad \times x/\sqrt{x^2+z^2} \Bigr) \\
    +& \boldsymbol{e}_y \Bigl( c_2 k A_0 \sin\theta_k J_0(\varkappa\sqrt{x^2+z^2})\mathrm{exp}[i k_z y + i\phi] \\
    &\; - c_1 i k A_0 \cos\theta_k J_1(\varkappa \sqrt{y^2+z^2})\mathrm{exp}[i k_z x] \\
    &\quad \times y/\sqrt{y^2+z^2} \Bigr) \\
    -& \boldsymbol{e}_z \Bigl( c_1 i k A_0 \cos\theta_k J_1(\varkappa \sqrt{y^2+z^2})\mathrm{exp}[i k_z x] z/\sqrt{y^2+z^2} \\
    &\, + c_2 i k A_0 \cos\theta_k J_1(\varkappa \sqrt{x^2+z^2})\mathrm{exp}[i k_z y + i\phi] \\
    &\quad \times z / \sqrt{x^2+z^2} \Bigr).
\end{split}
\label{eq:CombinedBField}
\end{equation}

\newpage
\bibliography{main}

\end{document}